\documentclass[usenatbib,usegraphicx]{mn2e}
\usepackage{subfigure}
\usepackage{longtable}
\usepackage{graphicx}
\usepackage{graphics}
\usepackage{keyval}
\usepackage{trig}
\usepackage{psfig,epsf}
\def\beq{\begin{equation}}
\def\eeq{\end{equation}}
\newcommand{\ltsima}{$\; \buildrel < \over \sim \;$}
\newcommand{\lesssim}{\lower.5ex\hbox{\ltsima}}

\title{Neutrinos as galactic dark matter in the Ursa Major galaxy group?}
\author[G. Gentile, H. S. Zhao and B. Famaey]{G. Gentile$^{1}$, H. S. Zhao$^{2}$ and B. Famaey$^{3}$\\
$^{1}$University of New Mexico, Department of Physics and Astronomy, 800 Yale Blvd NE, Albuquerque, New Mexico 87131, USA\\  
$^{2}$SUPA, School of Physics and Astronomy, University of St. Andrews, KY16 9SS Scotland\\
$^{3}$Institut d'Astronomie et d'Astrophysique, Universit\'e Libre  
de Bruxelles,  
CP 226, Bvd du Triomphe, B-1050, Bruxelles, Belgium\\
}

\begin{document}

\date{Accepted ... Received ... ; in original form ...}

\pagerange{\pageref{firstpage}--\pageref{lastpage}} \pubyear{2007}

\maketitle

\label{firstpage}

\begin{abstract}

We present the analysis of 23 published rotation curves of disk galaxies belonging to the Ursa Major 
group of galaxies, with kinematics free of irregularities. The rotation curves are analysed in the 
context of MOND (Modified Newtonian Dynamics). 
We add an extra component to the rotation curve fits, in addition to the stellar and 
gaseous disks: a 
speculative halo of constant density made of, e.g., neutrinos, which would solve the bulk of the problem currently
faced by MOND on rich galaxy clusters scales. We find that this additional unseen mass density is poorly constrained
(as expected a priori, given that a neutrino halo never dominates the kinematics), but
we also find that the best-fit value is non-zero: $\rho = 3.8 \times 10^{-27}$ g cm$^{-3}$, and that a 
zero-density is marginally excluded with 87\% confidence; also, the 95\% confidence upper limit for the density 
is $\rho = 9.6 \times 10^{-27}$ g cm$^{-3}$. These limits are slightly above the
expectations from the Tremaine-Gunn phase space constraints on ordinary 2~eV neutrinos, but in accordance with the maximum density expected for one or two species of 5~eV sterile neutrinos.
 
\end{abstract}


\section{Introduction}
\protect\label{sec:intr}
The current dominant paradigm in cosmology is that galactic dark matter is made of non-baryonic weakly-interacting massive particles, the so-called cold dark matter (CDM). However, on galaxy scales, observations are at variance with a sizeable list of CDM predictions (e.g., Klypin et al. 1999; Gentile et al. 2004, 2005, 2007a, 2007b, 2007c; Simon et al. 2003; Famaey \& Binney 2005; Kuzio de Naray et al. 2006), whilst one observes that the gravitational field is closely related to the baryonic distribution in spiral galaxies (e.g. McGaugh et al. 2007; Famaey et al. 2007a) through the modified Newtonian dynamics (MOND) relation of Milgrom (1983): in MOND, for gravitational accelerations $a$ 
below $a_0 \approx 10^{-8} {\rm cm} \, {\rm s}^{-2}$ the effective  
$a$ approaches $(g_{\rm N} a_0)^{1/2}$ where $g_{\rm N}$  
is the usual Newtonian gravitational field. This leads to remarkable fits of galactic rotation curves over five decades in mass ranging from tiny dwarfs (e.g., Gentile et al. 2007a, 2007c) through early-type spirals (e.g., Sanders \& Noordermeer 2007) to massive ellipticals (Milgrom \& Sanders 2003), without resorting to galactic dark matter. MOND also accounts very well for observed scaling relations of galaxies (e.g., Sanders \& McGaugh 2002; McGaugh 2004, 2005). Moreover, the mysteriously small but non-zero cosmological constant also finds a natural answer in some co-variant versions of MOND (e.g., Zhao 2007).

However, despite all these successes, MOND fails to completely explain away the mass discrepancy in clusters of galaxies (Aguirre, Schaye \& Quataert 2001; Sanders 2003, 2007; Pointecouteau \& Silk 2005; Angus, Famaey \& Buote 2007a): consequently, these objects still require substantial amounts of non-luminous matter in MOND. The recent weak-lensing observations of the ``bullet cluster" (Clowe et al. 2006, Angus et al. 2007b) have moreover constrained this non-luminous component to be collisionless, e.g. dark baryons in some non-dissipative form (MACHOs or small dense clumps of cold gas). Another possibility, suggested by Sanders (2003), is that ordinary neutrinos have a mass of the order of 2~eV (near the present experimental limit), and thus contribute significantly to the mass budget of galaxy clusters. 
These hot dark matter particles
are fermionic and have a maximum phase space density
set by equilibrium in the early Universe.
Reexpressing this phase space limit from Tremaine \& Gunn (1979),
and assuming 3 species of ordinary neutrinos of nearly equal mass,
Sanders (2003) noted that in a relaxed cluster of virialised gas temperature $T$, 
if the neutrino fluid has the same temperature as the gas, the maximum neutrino density is:

\begin{equation}
\rho_{\nu} = 4.8 \times 10^{-27} \left(\frac{m_{\nu}}{2 {\rm~ eV}}\right)^4 \left(\frac{T}{1 \rm{~keV}}\right)^{3/2}~{\rm g~ cm}^{-3},
\label{density}
\end{equation}

\noindent where $m_{\nu}$ is the ordinary neutrino mass.

It has indeed been shown that galaxy clusters scaling relations, including the luminosity-temperature relation, are then naturally reproduced in MOND (Sanders 2007), while such a non-baryonic component of moderately heavy neutrinos helps fitting the angular power spectrum of the Cosmic Microwave Background in relativistic MOND theories (e.g., Skordis et al. 2006). However, it has recently been shown that such ordinary 2~eV neutrinos would fail to explain the mass discrepancy in X-ray emitting groups of galaxies, where the X-ray temperature falls below 2 keV (Angus et al. 2007a), because the maximum neutrino density would be too low. This notion might be related to the fact that, due to their unrelaxed nature, the temperature of hot gas in groups is not representative of the actual temperature of the neutrino fluid. Alternatively, this could rather hint at the presence of extra baryonic mass in the form of cold dense clumps of cold gas (e.g. Pfenniger, Combes \& Martinet 1994; Pfenniger \& Combes 1994), or at the existence of one or two species of light sterile neutrinos with masses of a few eV (e.g. Maltoni et al. 2004, Maltoni \& Schwetz 2007).

In any case, if neutrinos do contribute significantly to the mass budget of rich galaxy clusters, it is then likely that, in the context of MOND, individual galaxies and galaxy groups possess extensive low-density neutrino halos. These were suggested by Sanders (2007) to possibly affect the weak lensing properties of galaxies on scales of hundreds of kpc. In this Letter, we rather investigate the possible contribution of neutrinos on the rotation curves of individual galaxies residing in groups in the context of MOND. To achieve this, we reanalyze the rotation curves in the Ursa Major galaxy group (Sanders \& Verheijen 1998) by adding as a new parameter a constant neutrino density.

\section{``MOND + neutrinos" models}
\label{mondneutrinos}

In MOND, the transition from the Newtonian regime (where $a = g_{\rm N}$) to the deep MOND regime (where $a = (g_{\rm N} a_0)^{1/2}$)
is regulated by an interpolation function $\mu(x)$, where $x = a/a_0$ and $a = g_{\rm N}/\mu(x)$. 
$\mu(x)$ should have the following asymptotic behaviours: $\mu(x) \rightarrow 1$ for $x >> 1$ and 
$\mu(x) \rightarrow a/a_0$ for $x << 1$. Traditionally the following functional form of $\mu(x)$ was used:
$\mu(x) = \frac{x}{(1+x^2)^{1/2}}$, but recently the preferred form has become the ``simple'' $\mu(x)$ 
proposed by Famaey \& Binney (2005):

\begin{equation}
\mu(x) = \frac{x} {1+x}
\end{equation}
 
This form of $\mu(x)$ has been shown to give disk galaxies' rotation curves fits which are as good as 
the standard $\mu(x)$, with realistic mass-to-light ($M/L$) ratios (Famaey et al. 2007a). Also,
contrary to the standard $\mu(x)$, it gives reasonable disk and bulge $M/L$ ratios in
early-type disk galaxies (Sanders \& Noordermeer 2007).

To motivate a 2 eV neutrino component in galaxies and galaxy groups, we note that
massive neutrinos are hot dark matter (decoupled from electrons and the
radiation at relativistic speed), but that these thermal relics
have presently cooled to very low speed. 
Indeed, since their momentum comparable to that of CMB photons today,
their speed is expected to be $\sim$ 30 km s$^{-1} \times (4/11)^{-1/3} =
21$ km s$^{-1}$. The factor $(4/11)^{-1/3}$ arises from entropy conservation
arguments in the early Universe (e.g. Weinberg 1972). 
They can however be heated up to 100-1000 km s$^{-1}$ speed by falling into the gravitational potentials of group and clusters of galaxies.

Eq. \ref{density} could be generalised to situations where the X-ray temperature
is unknown or unapplicable.  Adopting $T = 1 {\rm~ keV} (\frac{\sigma}{400 {\rm~ km~ s}^{-1}})^2$ from Sarazin
(1986), we can convert gas temperature to the 1-dimensional equilibrium
velocity dispersion $\sigma$.  The Tremaine-Gunn limit for three species of ordinary neutrinos (and their antiparticles) of nearly equal mass can then be recasted as:

\begin{eqnarray}
\nonumber
\left(\frac{\rho_{\nu}}{4.8 \times 10^{-27}~ {\rm g~ cm}^{-3}}\right)^{1/3} \left(\frac{m_{\nu}}{2 {\rm~ eV}}\right)^{-4/3} = \frac{\sigma}{400~ {\rm km~ s}^{-1}} \\
 = 
\left(\frac{T}{1 {\rm ~keV}}\right)^{1/2}
\end{eqnarray}
Note that $4.8 \times 10^{-27}~ {\rm g~ cm}^{-3} = 7 \times 10^{-5}~ {\rm M}_\odot~{\rm pc}^{-3}$.

\section{Data}

The Ursa Major group is an ideal example to perform this kind of rotation curve analysis. 
Galaxy groups are typically dominated by spiral galaxies for which the density profiles can be measured by analysing their rotation curves. Furthermore, such groups are usually diffuse and of relatively low mass. Consequently, the external field effect in galaxy groups is less important than in more massive galaxy clusters (Bekenstein \& Milgrom 1984, Famaey, Bruneton \& Zhao 2007b,
Angus \& McGaugh 2007, Wu et al. 2007). The Ursa Major galaxies do not form a rich, ``classical" virialised cluster, such as the ones for which 2~eV neutrinos have been invoked (Sanders 2003, 2007) as a solution to the MOND mass discrepancy problem. It is known for a while that such non X-ray emitting groups do not exhibit a mass discrepancy problem in MOND (e.g., Milgrom 2002), but within the uncertainties due notably to the unrelaxed nature of the group, a low density extra mass component cannot be excluded.

Note that an individual spiral galaxy is expected to be surrounded by a very shallow and extended neutrino halo (Sanders 2007), according to the phase-space contraint which corresponds to a temperature below 0.1 keV (i.e. typically $\sigma \lesssim 120$ km s$^{-1}$ for an individual galaxy, see e.g. Battaglia et al. 2005 for the Milky Way). Such halos, surrounding individual galaxies in a group, would merge with the halo of the group as a whole, for which the equivalent ``virialised gas temperature'' is T=0.14 keV, corresponding to a dispersion of ~150 km s$^{-1}$.
However, given that the Ursa Major group
is not virialised (Sanders \& Verheijen 1998), this value can only be taken as a very
rough estimate.

Hereafter, we model the neutrino component around each galaxy as the halo of the whole Ursa Major group, with an isotropic constant density.
 Given the uncertainties (the actual neutrino 
temperature, their density profile, the physical distances of the galaxies from the group
centre), it is a realistic assumption since a halo of ordinary 2~eV neutrinos is expected to feature a 
constant density core of several hundreds kpc (Sanders 2003). More detailed neutrino density distributions
will be the subject of future studies.
We then just fit rotation curves the usual way in MOND, but adding this extra component.

One of the great advantages of 
the galaxies of this sample is that they can be assumed to lie at approximately the same 
distance, since they belong to the same group. Contrary to Sanders \& Verheijen (1998),
who assumed a distance of 15.5 Mpc, here we took the distance to be 18.6 Mpc, the value 
calculated by Tully \& Pierce (2000) from the Tully-Fisher relation and 
Cepheid-determined distances to nearby galaxies;
we also note that Sakai et al. (2000) derive a slightly higher value of the distance,
20.7$\pm$3.2 Mpc, consistent with 18.6 Mpc.

The data that we used in the present paper were taken from Sanders \& Verheijen (1998). The authors 
present HI rotation curves of 30 galaxies belonging to the Ursa Major group. 
The derivation of the rotation curves is 
presented in Verheijen (1997) and Verheijen \& Sancisi (2001); they were derived from HI observations
performed at the WSRT (Westerbork Synthesis Radio Telescope). 
The original sample included 62 galaxies
brighter than $M_B \approx -16.5$, but 32 of them were not considered by Sanders \& Verheijen for their
kinematical analysis because they either had a too low inclination, they were lacking HI, they were 
interacting with neighbouring galaxies, or had too few independent points in the rotation curve.
Their final sample of 30 galaxies spans a large range of Hubble types. In the present
analysis, however, we only consider the 23 galaxies which are free of 
kinematical disturbances.
The stellar mass distribution is derived from NIR ($K'$-band) photometry (Tully et al. 1996), 
and the linear resolution
of the rotation curve is 0.75 kpc.

\section{Results and discussion}

We attempted to constrain the neutrino density, using the MOND hypothesis, by
fitting the 23 rotation curves discussed in the previous section. 

The rotation curves were fitted the usual way in MOND, but adding an extra component in the 
fits due to the putative neutrino halo. We used the simple $\mu(x)$
function (see Section \ref{mondneutrinos}). The global parameters of the fit
were $a_0$ and the additional (constant) density $\rho_\nu$: they both were constrained to be the same for all the galaxies
of the sample. 
We note that keeping $a_0$ as a free parameter can compensate any errors on the (fixed) distance.
Then, the stellar ($K'$-band) mass-to-light ($M/L$) ratio was left as an 
additional free parameter, separately for each galaxy. 
The uncertainties on the fitted parameters were derived from the $\chi^2$ statistics
of the global fit performed on the 23 rotation curves of the sample.

\begin{figure}   
\psfig{figure=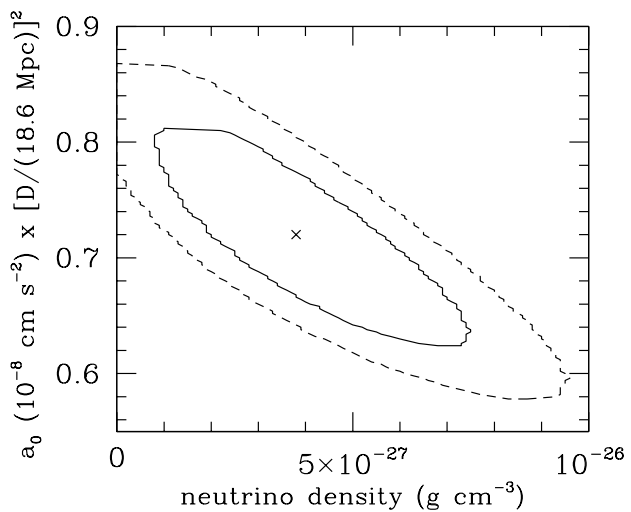,width=8.3cm,angle=0}   
\caption{
68\% (solid curve) and 95\% (dashed curve) confidence 
levels. D is the distance and 
the cross marks the best-fit value.
}
\label{contours}   
\end{figure}

We find a 1-$\sigma$ allowed range for $\rho_\nu$ between 
$0.8 \times 10^{-27}$ g cm$^{-3}$ and 
$7.4 \times 10^{-27}$ g cm$^{-3}$, the best-fit value being $\rho_{\nu} 
= 3.8 \times 10^{-27}$ g cm$^{-3}$. 
The corresponding best-fit value of $a_0$ is $0.72
\pm 0.09 \times 10^{-8}$ cm s$^{-2}$. In Fig. \ref{contours} we 
plot the confidence levels in parameter space: as shown by Begeman, Broeils
\& Sanders (1991), fits with variable $a_0$ and fixed distance $D$ are essentially identical
to fits with fixed $a_0$ and variable $D$, because the product $a_0 D^2$ comes in
the computation of the MOND circular velocity.  
Note that higher distances suggest a lower neutrino density.
Also, a negative neutrino density cannot be measured, thus any
fit of the density leads to a non-negative value.

\begin{figure}   
\psfig{figure=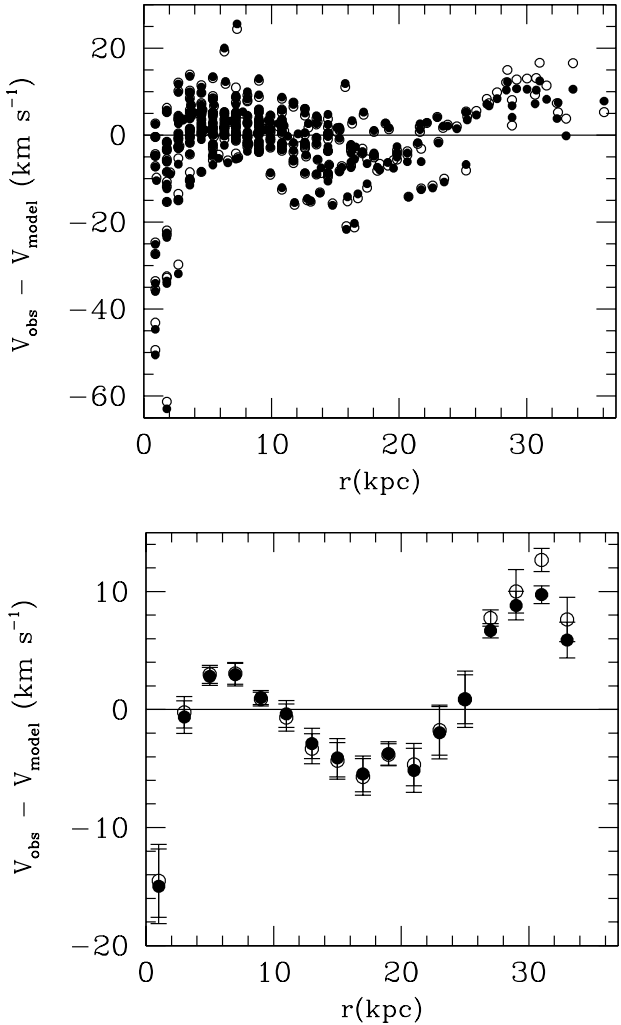,width=8.3cm,angle=0}   
\caption{
Residuals of the fits. Top: all the data. Bottom: data binned in radial bins of 2 kpc;
the errorbars are derived from the uncertainty on the mean in each bin.
Full circles represent the fits with neutrinos, empty circles
are the fits without neutrinos.
}
\label{resid}   
\end{figure}

As expected a priori, the constraints on $\rho_{\nu}$ are very weak, as the neutrino halo
never dominates the kinematics due to its very low density. 
Fig. \ref{resid} shows the overall effect of adding a neutrino component:
at large radii, where the neutrinos contribution is larger,
the residuals of the fits become smaller than without neutrinos. In the outer parts, 
both model rotation curves are below the observations, but the curves with neutrinos
are closer to the observations. 
The wave-like behaviour of the residuals in Fig. \ref{resid} will be studied 
in a future paper: different
$\mu$-functions might explain some systematic residual of the rotation curves 
(Sanders \& Verheijen 1998; Milgrom \& Sanders 2007).
However, using the standard $\mu$-function does not get rid of 
the wave-like behaviour of Fig. \ref{resid}.
We also note that $\rho_{\nu} = 0 $
is excluded with 87\% confidence. 

Let us now check whether the neutrino fits are consistent with the Tremaine-Gunn limit
and the global observation of Ursa Major.
Plugging the 1-$\sigma$ range of our rotation curve-determined value
($0.8 \times 10^{-27}$ to $7.4 \times 10^{-27}$) in the Tremaine-Gunn limit for ordinary 2~eV neutrinos, 
we predict  that $\sigma = 220 - 460 {\rm~ km~ s}^{-1}$, or equivalently that $T = 0.3 - 1.3 $ keV.
These should be the typical equilibrium
dispersion and temperature for characterizing
the overall potential which the 2~eV neutrinos are in.   
This equilibrium velocity dispersion is probably marginally consistent
with the global potential of the Ursa Major group:
no X-ray emission has been detected, but it is observed to have a galaxy
velocity dispersion of $\sigma_{obs} =148 \pm 6 {\rm ~km~ s}^{-1}$
(Tully 1987).  However, Ursa Major is
unrelaxed, so $\sigma_{\rm obs}$ is likely an under-estimate of the potential.
On the other hand, the lack of X-ray implies the intergalactic gas in Ursa is most certainly below
1 keV.  Adopting $\sigma=148$ km s$^{-1}$, the data is slightly pushing for a somewhat higher than 2 eV neutrino mass, which could be reminiscent of the fact that ordinary 2~eV neutrinos are not found to explain away the discrepancy in X-ray emitting groups (Angus et al. 2007a). 

Let us finally note that a possible effect on all these results is the external field effect (EFE), discussed in Section \ref{mondneutrinos}:
galaxies in clusters are embedded in an external field much higher than the typical value of 0.01$a_0$ in which 
the Milky Way is embedded (Famaey et al. 2007b, Wu et al. 2007),
so even when their internal acceleration is much smaller than $a_0$ they are actually not in a MOND regime.
However, as argued in Sanders \& Verheijen (1998), the EFE is expected to be of order 0.02$a_0$ in the Ursa Major group, 
meaning that we can neglect it as a first order approximation.

\section{Conclusions}

We analysed 23 rotation curves of disk galaxies with regular kinematics belonging to the Ursa Major 
group of galaxies, presented by Sanders \& Verheijen (1998). 
The rotation curves are fitted within the MOND (Modified Dynamics, Milgrom 1983) context, where
traditionally only the visible matter (gas and stars) is needed to fit rotation curves. 
However, given that MOND shows a mass discrepancy in 
rich galaxy clusters which can be mainly solved by adding a halo of 2~eV ordinary neutrinos (Sanders 2003), we add a third
component to the MOND fits, with the same constant density for all galaxies of the sample. It is however important to bear in mind that this extra mass component in our fits does not necessarily have to be ordinary neutrinos.

The fits give a rather unconstrained best-fit value of the additional constant density: $\rho_{\nu} = 3.8^{+3.6}_{-3.0} 
\times 10^{-27}$ g cm$^{-3}$. Such weak constraints
are to be expected since a neutrino halo is never the dominant component in the kinematics.
However, a zero additional density is marginally excluded, with 87\% confidence. These results
show that neutrino (or any other dark matter) effects are worth being modelled in MOND in the future, with better data and larger samples.

We however comment that there is a slight tension between the rotation curve-determined neutrino density and the Tremaine-Gunn limit for three species of ordinary 2~eV neutrinos, assuming that the velocity dispersion is a good estimator of the neutrino fluid temperature (which is unsure in such an unrelaxed cluster). Indeed, from Eq.(1), and for $T=0.14$~keV, the maximum neutrino density for UMa  should be $2.5 \times 10^{-28} (m_\nu/2{\rm eV})^4$ g cm$^{-3}$. For ordinary 2~eV neutrinos, this density is a factor 15 smaller than our best-fit value. A similar tension is observed in fitting weak-lensing data of high-z X-ray clusters (Takahashi \& Chiba 2007, Angus et al. 2007b), temperature profiles of X-ray emitting groups (Angus et al. 2007a), and velocity dispersions of galaxies at the center of  rich clusters (Richtler et al. 2007). This means that the narrow window of ordinary neutrino mass of 2-2.2 eV is presently tightly squeezed by opposite constraints from astronomical data and particle physics experiments. Actually, this tension could rather hint at the existence of one, two or three species of light sterile neutrinos (Maltoni \& Schwetz 2007).  For three species of sterile neutrinos with a 5~eV mass (and ordinary neutrinos with a negligible mass), a mass $m_\nu=5$~eV can be directly plugged in Eq.~(1). In the case of two species of sterile neutrinos, the maximum density must then be divided by a factor 1.5, and in the case of one species by a factor 3. This means that for two species of 5~eV sterile neutrinos, the maximum density for a temperature of 0.14~keV is  $6.5\times 10^{-27}$ g cm$^{-3}$, and for one species $3.25\times 10^{-27}$ g cm$^{-3}$, densities much closer to our rotation curve-determined values. These one or two species of 5~eV sterile neutrinos could additionally solve the problem of the MOND missing mass at the center of high-z X-ray clusters (Takahashi \& Chiba 2007, Angus et al. 2007b), in X-ray emitting groups (Angus et al. 2007a), and in individual galaxies at the center of clusters (Richtler et al. 2007).

Lastly, we note that doing similar rotation curve work in the much hotter
Virgo and Coma clusters might reveal a stronger effect of the neutrinos,
expected to be 100 times denser in Coma than in Ursa Major.  It is thus of interest to fundamental physics to push for accurate resolved kinematic data of galaxies in the Virgo and Coma clusters.

\section*{acknowledgements}
BF is a Research Associate of the FNRS.

\label{lastpage}

\end{document}